\documentclass{elsart}
\usepackage[dvips]{graphics,epsfig,rotating}
\usepackage{subfigure}
\usepackage{amstext}
\usepackage{amssymb}
\usepackage{amsmath}
\usepackage{float}
\usepackage{wrapfig}
\input{epsf}
\usepackage{array,booktabs,arydshln,xcolor}

\newcounter{bla}
\newenvironment{refnummer}{%
\list{[\arabic{bla}]}%
{\usecounter{bla}%
 \setlength{\itemindent}{0pt}%
 \setlength{\topsep}{0pt}%
 \setlength{\itemsep}{0pt}%
 \setlength{\labelsep}{2pt}%
 \setlength{\listparindent}{0pt}%
 \settowidth{\labelwidth}{[9]}%
 \setlength{\leftmargin}{\labelwidth}%
 \addtolength{\leftmargin}{\labelsep}%
 \setlength{\rightmargin}{0pt}}}
 {\endlist}
\newcommand{\n }{\\ \nonumber }
\newcommand{\pd }{Pad\'{e} }

\newcommand{\PD }{Pad\'{e}}
\newcommand{\la }{\leftarrow }
\newcommand{\q }{\quad }
\newcommand{\x }{\textbf }
\newcommand{\lm }{\lambda }

\begin{document}
\begin{frontmatter}

\title{Numerical Regge pole analysis of resonance structures in elastic, inelastic and reactive state-to-state integral cross sections}

\author{ E. Akhmatskaya$^{a,c}$, D. Sokolovski$^{b,c}$, and C. Echeverr\'ia-Arrondo$^{b}$}


\thanks[author]{dgsokol15@gmail.com}

\address[a]{Basque Center for Applied Mathematics (BCAM),\\ Alameda de Mazarredo, 14  48009 Bilbao, Bizkaia, Spain}

\address[b]{Department of Physical Chemistry, University of the Basque Country, Leioa, 48940, Spain}

\address[c]{IKERBASQUE, Basque Foundation for Science, E-48011 Bilbao, Spain}


\begin{abstract}
 We present a  detailed description of a  \texttt{FORTRAN} code for evaluation of the resonance contribution a Regge trajectory makes to the integral state-to-state cross section (ICS) within a specified range of energies. 
 The contribution is evaluated with the help of the Mulholland formula [Macek {\it et al} (2004)] and its variants  [Sokolovski {\it et al} (2007); Sokolovski and Akhmatskaya (2011)]. Regge pole positions and residues are obtained by analytically continuing $S$-matrix element, evaluated numerically for the physical values of the total angular momentum, into the complex angular momentum plane using the \texttt{PADE\_II}  program [Sokolovski  {\it et al} (2011)].
The code decomposes an elastic, inelastic, or reactive ICS into a structured, resonance, and a smooth, 'direct', components, and attributes  observed resonance structure to resonance Regge trajectories. The package has been successfully tested on several models, as well as the F + H$_2$$\to$ HF+H benchmark reaction. Several detailed examples are given in the text.

\begin{flushleft}

\end{flushleft}


\end{abstract}

\end{frontmatter}


{\bf PROGRAM SUMMARY}

\begin{small}
\noindent
{\em Manuscript Title:} Numerical Regge pole analysis of resonance structures in elastic, inelastic and reactive state-to-state integral cross sections.               \\
{\em Authors:} E. Akhmatskaya, D. Sokolovski,  and  C. Echeverr\'ia-Arrondo                 \\
{\em Program Title:} \texttt{ICS\_Regge}                      \\
{\em Journal Reference:}                                      \\
{\em Catalogue identifier:}                                   \\
{\em Licensing provisions:}   Free software license                                \\
{\em Programming language:} \texttt{FORTRAN 90}                        \\
{\em Computer:} Any computer equipped with a \texttt{FORTRAN 90}
compiler\\
{\em Operating system:} UNIX, LINUX                           \\
{\em RAM:} 256 Mb                                       \\
{\em Has the code been vectorised or parallelised:} no                                      \\
{\em Number of processors used:} one                          \\
{\em Supplementary material:} \texttt{PADE\_II}, \texttt{MPFUN} and \texttt{QUADPACK}  packages, validation suites, script files, input files,
readme files, Installation and User Guide
{\em Classification:} Molecular Collisions                    \\
{\em External routines/libraries:} none \\ 
{\em CPC Program Library subprograms used:}   N/A          \\
{\em Nature of problem:}\\
 The package extracts the positions and residues of resonance poles 
 from numerical scattering data supplied by the user. This information is then used for the analysis of resonance structures observed in elastic, inelastic and reactive integral cross sections.
 \\
{\em Solution method:}\\
 The $S$-matrix element is analytically continued in the complex plane 
 of either energy or angular momentum with the help of \pd 
 approximation of type II. Resonance  Regge
trajectories are identified and their 
contribution to the integral cross section are evaluated. 
   \\
{\em Restrictions:}\\
None.\\
  {\em Unusual features:}\\
  Use of multiple precision $MPFUN$ package.   \\
{\em Additional comments:} none\\
{\em Running time:}\\
from several minutes to hours depending on the number of energies involved, the  precision level chosen and the number of iterations performed.\\
{\em References:}
\begin{refnummer}
\item J. H. Macek, P. S. Krstic, and S. Yu. Ovchinnikov, 
Phys. Rev. Lett. {\bf 93},  (2004) 183203.
\item D. Sokolovski, D.De Fazio, S.Cavalli and V.Aquilanti,
J. Chem. Phys., {\bf 126} (2007) 12110.
\item D. Sokolovski and E. Akhmatskaya, Phys. Lett. A, {\bf 375},  (2011) 3062. 
\item D. Sokolovski, E. Akhmatskaya and S. K. Sen, Comp. Phys. Comm. A, {\bf 182},  (2011) 448.
\end{refnummer}

\end{small}

\newpage


\hspace{1pc}
{\bf LONG WRITE-UP}

\section{Introduction}

In the last fifteen years the progress in crossed beams experimental techniques has been matched by the  development  of state-of-the-art computer codes capable of modelling atom-diatom elastic, inelastic and reactive differential and  integral cross sections (ICS) \cite{CODE1}- \cite{CODE2}. The ICS, accessible to measurements in crossed beams, are often structured, and so offer a large amount of useful information about details of the collision or reaction mechanism. This information needs to be extracted and analysed, which often presents a challenging task.
One distinguishes two main types of collisions: in a {\it direct} collision the partners depart soon after the first encounter, while in a  {\it resonance} collision they form an intermediate complex 
(quasi-molecule) which then breaks up into products (reactive case) or back into reactants (elastic or inelastic case). The resonance pathways may become important or even dominant at low collision energies. For this reason, accurate modelling and understanding of resonance effects gains importance in such fields as cold atom physics and chemistry of the early universe. 
\newline
Once the high quality scattered matrix is obtained numerically, one needs to understand the physics of the reaction, often not revealed until an additional analysis is carried out.
In particular, resonances invariably leave their signatures on the integral state-to-state cross sections.
In this paper we propose and describe software for the analysis of such resonance patterns.
Relevant information on the Regge poles can be found in Refs.\cite{REGGE1}-\cite{REGGE2}. Some applications of the poles to the angular scattering and integral cross sections are discussed in Refs.\cite{DCS1}-\cite{DCS2} and \cite{INT1}-\cite{INT2}, respectively.
For a description of the type-II \PD\q approximation, used by the software, the reader is referred to Refs.\cite{PADE1}-\cite{PADE2}. 
\section{Background and theory}
We start with a brief review of the techniques required for our analysis.
 \subsection{The integral, or total, scattering cross section}
An integral (total) cross section, $\sigma(E)$, gives the total number of incident particles with the energy $E$, scattered in all possible directions per unit time, for unit incoming flux.
This definition is valid for a single particle scattered by a central force, as well as for an atom $A$ colliding with a diatomic molecule ($BC$). 
\newline
In the latter case, the collision partners may part in the same arrangement ($A+BC$), with the molecule in the same internal state described by the vibrational ($v$), rotational ($j$), and helicity ($\Omega$ = projection of $j$ onto the final atom-diatom velocity) quantum numbers. 
This is {\it elastic} scattering. 
\newline
In an {\it inelastic} scattering event, the arrangement ($A+BC$) remains the same, 
yet the internal quantum numbers of the molecule $BC$ are changed, ($v,j,\Omega \to v',j',\Omega')$. 
\newline
Finally, in a {\it reactive} event the atom $B$ becomes attached to the atom $A$ $(A+BC\to AB+C)$, and the quantum numbers of the newly formed diatomic may take any values allowed by the conservation laws.
\newline
The probability amplitude for each process is defined for an energy  $E$ ({\it total}, if it includes the internal energy of the reactant diatomic, or {\it collision} $E^{coll}$, if it does not), and a value of the total angular momentum $J$ (we use $\hbar=1$). It is given by complex-valued scattering ($S$-) matrix element $S^{\alpha}_{\nu' \la \nu}(E,J)$, where $\alpha = elastic, inelastic, reactive$, and $\nu$ is the shorthand for $(v,j,\Omega)$.
\newline
In all cases, the integral cross section can be written as a partial wave sum (PWS) over all physical (i.e., integer) values of the total angular momentum. For the elastic channel,  where the interference with the incoming wave must be taken into account, one has 
\begin{eqnarray}\label{1}
 \sigma^\alpha_{\nu^{\prime} \gets \nu}(E) = 
 \frac{2\pi}{k_{\nu}^{2}}\sum_{J=J_{min}}^{\infty}(J+1/2)
 |S^\alpha_{\nu \gets \nu}(E,J)-1|^2,\q \alpha = elastic,\\ \nonumber
\end{eqnarray}
where
\begin{eqnarray}\label{1a}
 J_{min}=|\Omega|, 
\end{eqnarray}
since the projection of the molecule's angular momentum on the relative velocity cannot exceed the total angular momentum $J$.
\newline
For an inelastic or reactive process there is no such interference, and the ICS takes the form
\begin{equation}\label{2}
 \sigma^\alpha_{\nu^{\prime} \gets \nu}(E) = 
  \frac{2\pi}{k_{\nu}^{2}}\sum_{J=J_{min}}^{\infty}(J+1/2)
 |S^\alpha_{\nu^{\prime} \gets \nu}(E,J)|^2,\q \alpha = inelastic\q or\q reactive,
\end{equation}
with 
\begin{eqnarray}\label{2a}
 J_{min}=max(|\Omega|,|\Omega'|). 
\end{eqnarray}
 In Eqs. (\ref{1}) and (\ref{2}) $k_\nu=\sqrt{2\mu E^{coll}}$ is the reactant's relative wave vector, which in the case of the single-channel potential scattering becomes the wave vector of the incoming plane wave, $k=\sqrt{2\mu E}$, $\mu$ being the corresponding reduced mass.
\newline
Identification and quantitative analysis of structures produced in the ICS by the capture of collision partners into long lived metastable states
 is the main subject of this paper. Despite their simplicity, Eqs.(\ref{1})-(\ref{2}) are not best suited for the task, since the $S$-matrix elements contain contributions from both the direct and the resonance mechanisms.  A more convenient representation, designed to separate the two contributions, was proposed by Macek {\it et al}  in their study of the proton scattering by hydrogen \cite{INT1}, and later used by Ovchinnikov {\it et al} who considered proton impact on  inert gas atoms \cite{ICS2}. The method relies on the concept of Regge poles.
 which we will discuss here only briefly.
 \subsection {Regge poles and Regge trajectories} 
 
Regge poles are the poles of $S$-matrix element $S_{\nu' \leftarrow \nu}(E,J)$ in the first quadrant complex plane of  the (continuous) variable $J$, evaluated at a fixed energy $E$ \cite{REGGE1}.
Sharp long-lived resonances manifest themselves as poles at $J=J_n(E)$, $n=1,2,...$, close to the real axis,
\begin{equation}\label{3}
S_{\nu' \leftarrow \nu}(E,J_n)=\infty, \quad Re J_n>0, \quad Im J_n>0.
\end{equation}%
Resonance Regge poles are closely related to the resonance {\it complex energy poles} \cite{FH3}, \cite{IH1}, \cite{ICS3}, but are more convenient for the analysis of the quantities given by partial wave sums at a fixed energy, like the ICS in Eqs. (\ref{1})-(\ref{2}). A Regge pole cannot disappear suddenly (except by coalescing with a complex zero of the $S$) and, as the energy changes, traces a continuous curve in the complex angular momentum (CAM) plane. These curves are known as {\it Regge trajectories}.
\newline
It is useful to distinguish between at least two types of Regge trajectories \cite{PLA}: those which at zero angular momentum correspond to a bound state supported by the potential ({\it type I}), and those
corresponding at $J=0$ to a potential's metastable state ({\it type II}). Regge trajectories [curves 
$Im J_n(E)$ vs. $Re J_n(E)$] of the type $(I)$ typically start, at low energies, on the real $J$-axis, and then move into the complex $J$-plane in its first quadrant. The trajectories of the type $(II)$
initially descend on the real $J$-axis, and affect an ICS in a different way \cite{PLA}. Our test cases, provided below, contain examples of each type.
\newline
 \subsection{The Mulholland formula}
Macek {\it et al} \cite{INT1}, \cite{ICS2} replaced the discrete sum over $J$ in Eq. (\ref{1}) by an integral, and then deformed the contour of integration to run along 
the imaginary $J$-axis, thereby picking contributions from the Regge poles in the first quadrant. The result is the decomposition 
\begin{eqnarray}\label{4}
 \sigma^{elastic}(E)=\frac{2\pi}{k^2}\int_0^\infty f(E,\lambda)\lambda d\lambda-Re \int_0^{i\infty}\frac{4\pi f(E,\lambda)/k^2}{1+\exp(-2i\pi\lambda)} \lambda d \lambda
\n
 -\frac{2\pi}{k^2}Re \sum_n\frac{4\pi i Res[f(E,\lambda_n)]\lambda_n}{1+\exp(-2i\pi\lambda_n)}
\end{eqnarray}
where  $f(E,\lambda)\equiv|S(E,\lambda)-1|^2$, $\lambda \equiv J+1/2$, $\lambda_n=J_n+1/2$ is the position of the $n$-th pole in the CAM plane, and $Res_n[f(E,\lambda_n)]$ is the residue at the $n$-th pole,
\begin{equation}\label{5}
Res[f(E,\lambda_n)]\equiv\lim_{\lambda \to \lambda_n}(\lambda-\lambda_n)f(E,\lambda).
\end{equation}
The integrals along the real and imaginary axis  are expected to be slowly varying function of the energy, with the resonance structure given mostly by the last sum in Eq.(\ref{4}).
The authors of Refs. (\ref{1})-(\ref{2}) traced their technique back to a 1928 paper by H. P. Mulholland 
\cite{Mull} (whose subject, one must say, had very little to do with the present analysis), 
and we will follow them in calling the result (\ref{4}) the  {\it Mulholland formula}.
\newline
The practical use of the formula (\ref{4}) is quite simple: one chooses a particular, say the $n$-th,  Regge trajectory, evaluates along it the contribution to the ICS given by the $N$-th term in the sum, 
\begin{eqnarray}\label{5a}
I^{Mull}_n(E)=-\frac{2\pi}{k^2}Re\frac{4\pi iRes[f(E,\lambda_n)]\lambda_n}{1+\exp(-2i\pi\lambda_n)},
\end{eqnarray}
 and then subtracts the result from the full ICS. If what is left is smooth and structureless, one can identify the pattern in the full ICS as arising from this particular resonance trajectory. There are many possible patterns. From (\ref{5a}) one readily sees that a pole contribution is greatly enhanced whenever the trajectory passes in the vicinity of a positive integer, $J_n\approx N > 0$  \cite{INT1}. This is hardly surprising since one recalls that  a true bound state requires an integer value of $J$, and that a sharp resonance is in fact  a bound state which is weekly connected to a continuum. Depending on the potential, a trajectory may follow in the vicinity of the real $J$-axis, passing by several integer $J$'s, and thus producing a series of peaks in the ICS \cite{PLA}, \cite{SE2}.  Or it can veer steeply deep into the CAM plane thus producing just one peak, or even no peaks at all  \cite{INT1}.  A trajectory following the real axis at a larger distance would produce in the ICS not peaks, but 
sinusoidal oscillations \cite{FH2}. A trajectory may start on the real axis, and then move deeper into the CAM plane, or descend to the real $J$-axis, and then follow it for some time. All these possibilities account a large variety of ways in which a resonance may produce a structure in an ICS.

\subsection{ Modifications of the Mulholland formula}
The Mulholland decomposition of the ICS (\ref{4}) is easily modified to the case 
of reactive and inelastic transitions as \cite{FH2} (below  $\alpha = inelastic$  or $reactive$)
\begin{eqnarray}\label{6}
 \sigma^{\alpha}_{\nu^{\prime} \gets \nu}(E) = \frac{2\pi}{k_\nu^2}\int_{J_{min}-1/2}^\infty|S_{\nu' \leftarrow \nu}(E,\lambda)|^2 \lambda d\lambda +\q\q\q\q\q
 \q\q\q\q\q\q\q\
 \n
 \frac{8\pi^2}{k_\nu^2}Im \sum_n\frac{\lambda_n Res[S_{\nu' \leftarrow \nu}(E,\lambda_n)]S^*_{\nu' \leftarrow \nu}(E,\lambda_n^*)}{1+\exp(-2i\pi\lambda_n)} +I_{\nu' \leftarrow \nu}(E), \q\q\q\q\q\q\q\q
\end{eqnarray}
where  $^*$ denotes complex conjugation, 
and $I_{\nu' \leftarrow \nu}$ contains integrals along the imaginary $\lambda$-axis, similar to the second term in Eq. (\ref{4}). In the following we will not be interested in the explicit  form of $I_{\nu' \leftarrow \nu}$. Equation (\ref{6}) is exact, and if necessary, $I_{\nu' \leftarrow \nu}$ can be evaluated by subtracting from $\sigma^{\alpha}_{\nu^{\prime} \gets \nu}(E)$, computed as a PWS, the first two terms in its r.h.s.
\newline
For an elastic transition in a multi-channel case,  $\alpha=elastic$, $\nu=\nu'$,
one replaces in Eq.(\ref{6}) 
\begin{equation}\label{7}
S_{\nu' \leftarrow \nu}(E,\lambda) \to S_{\nu' \leftarrow \nu}(E,\lambda)-1.
\end{equation}
The formula (\ref{6}), which reduces to (\ref{4}) in the single-channel case, 
is to be used just as described at the end of the previous Subsection, and is employed in our computer code.
Previously, it has been applied to analyse resonance structures in the ICS of the benchmark $F+H_2\to HF+H$ reaction \cite{FH2}, \cite{ICS3}.
\newline
Returning to the single-channel case, one notes \cite{PLA}, \cite{SE2} that the factor 
\begin{equation}\label{8}
[1+\exp(-2i\pi\lambda_n)]^{-1}=-\sum_{M=1}^{\infty}\exp(2\pi i M J_n)
\end{equation}
is a sum of a geometric progression in which the first term, a unity, is missing. The physical picture is that of 
an intermediate metastable complex which rotates and decays at the same time.
The optical theorem \cite{SE2} allows one to interpret the terms in the r.h.s of Eq.(\ref{8}) as the contributions a rotating complex makes to the forward scattering amplitude after $M=1,2...$ complete rotations. The missing term in (\ref{8}) is just what the complex contributes to 
forward scattering immediately after its formation, before  completing even one rotation. Thus, the $M=0$ term can be subtracted from the first, and added to the second term in Eq.(\ref{6}). This yields a modified Mulholland formula \cite{PLA}, 
\begin{eqnarray}\label{9}
 \sigma
 (E) = \frac{2\pi}{k^2}\int_\Gamma|S
 (E,\lambda)-1|^2 \lambda d\lambda +
 \frac{8\pi^2}{k^2}Im \sum_n\frac{\lambda_n Res[S
 (E,\lambda_n)]}{1+\exp(2i\pi\lambda_n)} 
 \n +I(E), \q
\end{eqnarray}
where the integration contour $\Gamma$ runs in the first quadrant of the CAM above all poles
which make significant Mulholland contributions to the ICS. Decomposition  (\ref{9}) was shown to achieve a better separation of the ICS into direct and resonance parts for a simple single-channel model in \cite{PLA}, and we include the modified formula in the present code. However, we have yet been unable to extend its application to the multichannel case, and the option for using (\ref{9}) remains open to single-channel potential scattering only.
\subsection{ \pd reconstruction of the scattering matrix element}
The application of the decompositions  (\ref{6})-(\ref{9}) requires the knowledge of the pole positions $\lambda_n=J_n+1/2$, and the corresponding residues, $Res[S(E,\lambda_n)]$.  Since the analytic continuation of $|f(E,\lambda)|^2$ from the real axis into the complex $\lambda$-plane is  $f(E,\lambda)f^*(E,\lambda^*)$, one also requires the values of $S^*(E,\lambda_n^*)$. These data are not usually available since a computer code used in modelling  a chemical reaction typically evaluates the $S$-matrix elements, $S(E,J)$, for the physical integer values of $J=0,1,2,...N$, with $N$ sufficiently large to converge the partial waves sums (\ref{1}) and (\ref{2}). Using these values, we construct a rational \pd approximant, ($[x]$ stands for the integer part of $x$)
\newline
\begin{eqnarray}\label{10}
  S_{\nu' \gets \nu}^{Pade}(E,J)\equiv
 K_{N} \textrm{exp}[i(aJ^{2}+ bJ+c)]
 \times \frac{\prod_{i=1}^{[N/2]}(J-Z_i)}
 {\prod_{i=1}^{[(N-1)/2]}(J-P_{i})},
 \end{eqnarray} 
 where $P_i(E)$ and $Z_i(E)$ stand for poles and zeroes of the approximant, respectively, and $K_N(A)$, $a(E)$, $b(E)$ and $c(E)$ are energy dependent constants.
The approximant is conditioned to coincide with $S(E,J)$ at the $N$ integer values of $J$,
\begin{eqnarray}\label{11}
 S_{\nu' \gets \nu}^{Pade}(E,J)=S_{\nu' \gets \nu}(E,J), \quad J=0,1,...N,
\end{eqnarray}%
and provides an analytic continuation of the exact function $S(E,J)$ in a region of the complex $J$-plane, containing the values (\ref{11}). Inside this region, the poles and zeroes of the approximant coincide with the true Regge poles and zeroes of the $S$-matrix element. 
The remaining poles and zeroes tend to mark the border of the region, beyond which \pd approximation fails \cite{G1}-\cite{PADE2}.
Thus, for a given pole $J_n=P_j$ we evaluate the required quantities as
\begin{eqnarray}\label{12}
Res[S_{\nu' \gets \nu}(E,\lambda_n)] =  K_{N} \textrm{exp}[i(aP_j^{2}+ bP_j+c)]
 \times \frac{\prod_{i=1}^{[N/2]}(P_j-Z_i)}
 {\prod_{i=1,i\ne j}^{[(N-1)/2]}(P_j-P_{i})},
 \end{eqnarray} 
and 
\begin{eqnarray}\label{13}
S^*(E,\lambda_n^*) =  K^*_{N} \textrm{exp}[-i(aP_j^{2}+ bP_j+c)]
 \times \frac{\prod_{i=1}^{[N/2]}(P_j-Z^*_i)}
 {\prod_{i=1}^{[(N-1)/2]}(P_j-P^*_{i})},
 \end{eqnarray} 
which provides the data needed for evaluating the resonance contributions in the Mulholland formulas (\ref{6}) and (\ref{9}).
\newline
The program used to construct the \pd approximant (\ref{10}) is essentially the  $\texttt{PADE\_II}$ code reported in the Ref. \cite{PADE2}. The minor modifications made to adapt it to the problem in hand will be described below.
\subsection{ A brief summary}
Our analysis consists in attributing to particular Regge trajectories resonance structures observed in an elastic, inelastic of reactive ICS obtained by numerical modelling of a scattering process. The behaviour of the scattering matrix element in the complex $J$-plane, required for evaluating the contribution a trajectory makes to the ICS, is reconstructed by using \pd approximants of type (II) \cite{PADE2}. The method gives a clear picture of how a resonance affects the ICS in a range of energies. It does not, however, reveal the physical origin of the resonance (e.g., its location in the entrance or exit channel on the potential surface), which must be established independently.
\section {ICS\_Regge package: Overview} 
\subsection{ Installation}
This version of \texttt{ICS\_Regge} is intended for IA32 / IA64 systems running the Linux operating system. It requires \texttt{Fortran} and \texttt{C} compilers. 
The software is distributed in the form of a gzipÕed tar file, which contains the  \texttt{ICS\_Regge} source code, \texttt{PADE\_II 1.1} source code, \texttt{QUADPACK} source code, test suites for each package, as well as scripts for running and testing the code. The detailed structures of each subpackage, \texttt{ICS\_Regge}, \texttt{PADE\_II 1.1} and \texttt{QUADPACK}, are presented in the Appendices C, D and E.  
For usersÕ benefits we supply a file \texttt{README} for each package in directories  \texttt{ICS}, \texttt{ICS/PADE} and \texttt{ICS/QUADPACK}. The files provide a brief summary on the code structure and basic instructions for users. The  \texttt{ICS\_Regge} Manual, \texttt{ICSManual.pdf}, is located in \texttt{ICS/} directory whereas \texttt{PADE\_II} Manual, \texttt{FManual.pdf}, can be found in \texttt{ICS/PADE} directory.
In addition, the documents describing \texttt{PADE\_II} and \texttt{QUADPACK} are available from \texttt{http://www.cpc.cs.qub.ac.uk/} (Catalogue identifier: \texttt{AEHO\_v1\_0)} and  
\newline
\texttt{http://www.netlib.org/quadpack/} respectively. 
Once the package is unpacked the installation should be done in the following order:

1.	Installing \texttt{QAUDPACK} library

2.	Installation of the \texttt{PADE\_II} package

3.	Installation of the \texttt{ICS\_Regge} package.  

Installation procedure for each subpackage is straightforward and can be successfully performed by following the instructions in the \texttt{README} files or / and in  \texttt{ICS\_Regge}, \texttt{PADE\_II} Manuals. Here we just want to notice that two ways of building the \texttt{ICS\_Regge} and \texttt{PADE\_II} executables are available. One is manual and is recommended for the first time users at the installation stage, whereas the fully automated way is useful at late stages of using the package. 
While a manual building assumes using a \texttt{Makefile\_UNIX} with the adjusted environmental variables, an automated building relies on the run script \texttt{runICS} with the built-in instructions for making the packages. The appropriate values of environmental variables are passed, in this case, through the input file \texttt{ICS/input/INPUT} controlled by a user (see section 5 for details).

\subsection {Testing}
Three test suits are prepared for each subpackage to validate the installation procedure. 

\subsubsection {Running the \texttt{QUADPACK} test}
To test the \texttt{QUADPACK} library, one has to run \texttt{quadpack\_prb.sh} script in \texttt{/QUADPACK/scripts} directory. 

The results of 15 tests can be found in \texttt{quadpack\_prb\_output.txt} file in 
\texttt{ICS/QUADPACK/test} directory. The message 
\newline
           \texttt{QUADPACK\_PRB: Normal end of execution}.
\newline
confirms that the code passed the validation test. 
\subsubsection {Running the \texttt{PADE\_II} test suite}
The input for 4 jobs,  \texttt{test1},  \texttt{test2},  \texttt{test3} and  \texttt{test4}, are provided in directories  \texttt{ICS/PADE/test/input/test\_name}, where \texttt{test\_name} is either  \texttt{test1} or  \texttt{test2} or  \texttt{test3} or  \texttt{test4}.
To submit and run a test suite, first, one can change the definition of the \texttt{Fortran} compiler, \texttt{F77}, \texttt{C} compiler, \texttt{CC}, their paths,  \texttt{COMP\_PATH} and  \texttt{CCOMP\_PATH} respectively, if necessary in \texttt{Makefile\_test}  in directory \texttt{ICS/PADE/src}, and then go to the directory \texttt{ICS/PADE/test} and run a script  \texttt{./run\_TEST}.
All tests will be run in separate directories. Each test takes about 1 - 5 minutes to run on a reasonably modern computer.
The message
\newline
\texttt{Test test\_name was successful}
\newline
appearing on the screen at the conclusion of the testing process confirms that the code passed the validation test \texttt{test\_name} . The results of the simulation can be viewed in the  \texttt{ICS/PADE/test/output/test\_name} directory.  
The message
\newline
 \texttt{Your output differs from the baseline!}
\newline
means that the calculated data significantly differ from that in the baselines. The files \texttt{PADE/test/output/test\_name/diff\_file} can be checked  to judge the differences. 
\subsubsection {Running the \texttt{ICS\_Regge} test suite}
For simplicity, the test suite for  \texttt{ICS\_Regge} is designed in the similar manner as the test suite for  \texttt{PADE\_II}. 
The input for 3 jobs,  \texttt{test1},  \texttt{test2} and  \texttt{test3} are provided in directories \texttt{ICS/test/input/test\_name} respectively, where  \texttt{test\_name} is either  \texttt{test1} or  \texttt{test2} or  \texttt{test3}. The definitions of the \texttt{FORTRAN} and \texttt{C} compilers and their paths, as well as the path of the  \texttt{PADE\_II} directory, can be changed in the \texttt{INPUT} files in each directory  \texttt{test\_name}, if required. 
The test suit can be run in the directory \texttt{ICS/test} using the following commands:
 \texttt{./run\_TEST < test\_data.txt}.
All tests are run in separate directories, \texttt{ICS/test/output/test1}, \texttt{ICS/test/output/test2} and \texttt{ICS/test/output/test3}.
Each test takes about 1 - 6 minutes to run on a reasonably modern computer.
To analyse the test results, the message on the screen at the conclusion of the testing process has to be inspected. 
The message
\texttt{ Test test\_name was successful}
confirms that the code passed the validation test  \texttt{test\_name}. The results of the simulation can be viewed in the \texttt{ICS/test/output/test\_name} directory.  
The message

        \texttt{test\_name  output differs from the baseline!}

        \texttt{Check your output in output/test\_name /ics.mull}

        \texttt{Baseline file: baselines/test\_name /ics.mull}

        \texttt{Differences: output/test\_name /error\_file}

means that the calculated data significantly differ from that in the baselines. The differences can be found in the files  \texttt{ICS/test/output/test\_name/error\_file}.
%
\section {Computational modules of ICS\_Regge} 
\subsection {\pd reconstruction and $\texttt{PADE\_II}$ options}
As discussed above, an important part of the calculation consists in performing an analytical continuation of the $S$-matrix element into the CAM plane.
The user has some flexibility in doing so. It concerns mostly the quadratic phase in Eq. (\ref{10}), which must itself be determined in the course of the \pd reconstruction. The need for separating this rapidly oscillating term arises from the fact that the \pd technique used here, works best for slowly varying functions. Thus, by removing the oscillatory component, one expands  the region of validity of the approximant in the CAM plane, which allows to correctly describe a larger number of poles.
The extraction of the quadratic phase proceeds iteratively. Since sharp structures in the phase of the $S$-matrix element usually come from the Regge poles and zeroes located close to the real axis, one defines a strip $\texttt{-dxl}<Im J<\texttt{dxl}$ around the real $J$-axis, and removes from the previously constructed approximant all poles and zeroes inside the strip. The smoother phase of the remainder is fitted to a quadratic polynomial, and this quadratic phase is then subtracted from the phase of the input values of $S$, after which a new approximant is constructed with these modified input data. The process is repeated \texttt{niter} times resulting in a (hopefully) improved \pd approximant.
There is no rigorous estimate of the improvement achieved, and the practice shows that in many cases using \texttt{niter  > 1} gives tangible benefits, while in some cases better results are achieved with \texttt{niter}=$1$ or $2$. It is for the user to decide on the best values of \texttt{dxl} and \texttt{niter} for a particular problem.
\newline
The input files required for running the \texttt{\PD\_II} code are stored in the input directory where they are labelled $1,2...N_E$.
A typical input file is given in Appendix B. The file differs from the similar input used in  \texttt{\PD\_II} package reported in \cite{PADE2} by one line added at the end, which should contain the collision energy in $meV$. 
\newline
Other parameters for \pd reconstruction are read from the \texttt{input/INPUT} file. These include the pathways to the \texttt{PADE} directory, to the \texttt{FORTRAN} and \texttt{C} compilers, and the compiler options (see \#16-\#21). 
\newline
The next entry (\#22) determines whether there should be a change of parity from the original data. This depends on the convention used in calculating the $S$-matrix elements, as explained in \cite{PADE2} (we use $0$ for \texttt{no} and $1$ for \texttt{yes}). 
\newline
Entry (\#23) decides whether 
one should remove the guessed values of the quadratic phase prior to the construction of the first \pd approximant in a series of iterations.  Its recommended value is $1$ (\texttt{yes}). 
\newline
Entry (\#24) determines if multiple precision routines should be used in calculating the \pd approximant. The recommended value is $1$ if the number of partial waves (PW) exceeds $40$.
It can also be used for a smaller number of (PW) to check the stability of calculations.
\newline
Entries (\#25 and  \#27 ) allow us to repeat the calculations with added non-analytical noise of magnitude \texttt{fac}, \texttt{nstime} times. This may be needed to check the sensibility of calculations to numerical noise. The recommended initial values are \texttt{nstime=0} and \texttt{fac=0}, in which case no noise is added.
\newline
Entry (\#26) determines the number of points in the graphical output from \texttt{\PD\_II} \cite{PADE2}.
\newline
Finally, the user has the options of changing  the number of partial waves, the number of iterations and the value of \texttt{dxl} for all files used in the current run by setting to $1$ \texttt{iover1}, \texttt{iover2} and \texttt{iover3} in entries (\#28, \#29 and  \#30). The corresponding parameters are reset to the values \texttt{nread1}, \texttt{niter1}, and \texttt{dxl1}, specified in the entries (\#31, \#32 and  \#33), respectively.

\subsubsection{Changes made to $\texttt{PADE\_II}$ }
The changes from the previous version \cite{PADE2} include 
\newline
(I) replacement of all the Numerical Algorithms Group (NAG) routines with ones available in the public domain, and 
\newline
(II) 
provision of additional controls allowing to change the parameters of \pd reconstruction for all energies in the run at once, without changing individual input files labelled $1$,$2$,..., as discussed in the previous Section. 
\newline
A brief summary of the changes made to subroutines is given below.
\newline
{\it Subroutine \texttt{FIT} (\texttt{fit.f})}
\newline
The NAG routine \texttt{g05ccf} has been replaced by a subroutine \texttt{svdfit} described in section 15x.4 of {\it Numerical Recipes in C: The Art of Scientific Computing (Second Edition), published by Cambridge.}
\newline
\newline
{\it Subroutine \texttt{ZSRND} (\texttt{zsrnd.f})}
\newline
The NAG routine \texttt{E02ACF}  has been replaced by a sequence of calls to the system routines \texttt{srand48}  and \texttt{drand48}.
\newline
\newline
{\it Subroutine \texttt{IDET\_SEED} (\texttt{idet\_seed.c})}
\newline
Added new routine generating the seed for \texttt{srand48}.
\newline
\newline
{\it Wrapper (\texttt{wrapper.c})}
\newline
Added wrapper allowing for calling C-routines \texttt{idet\_seed.c}, \texttt{srand48} and \texttt{drand48} in the \texttt{Fortran} code.
{\subsection{The structure of $ \texttt{ICS\_Regge}$ } 

The $ \texttt{ICS\_Regge}$ application is a sequence of 12 FORTRAN files. The files are listed below, and their functions are explained.
\newline
\newline
{\it Program \texttt{ICS\_Regge} (\texttt{ICS\_Regge.f})}
\newline
Main program.
\newline
\newline
{\it Subroutine \texttt{READ1} (\texttt{read1.f})}
\newline
Opens and reads the original input file at a given energy.
\newline
\newline
{\it Subroutine \texttt{READ} (\texttt{read.f})}
\newline
Reads the parameters of the \pd  reconstruction at a given energy.
\newline
\newline
{\it Subroutine \texttt{SORT} (\texttt{sort.f})}
\newline
At a given energy, selects poles and zeroes in the specified region of the complex angular momentum ($J$-) plane and discards pole/zero pairs (Froissart doubles).
\newline
\newline
{\it Subroutine \texttt{TCROSS} (\texttt{tcross.f})}
\newline
Evaluates the partial wave sum (PWS) for the integral cross section using the original data and also the \pd approximant  for integer values of  $J$.  Estimates the error of the \pd reconstruction.
\newline
\newline
{\it Subroutine \texttt{TCINTE} (\texttt{tcinte.f})}
\newline
Replaces the discrete PWS by integration over continuous values of J. Evaluates the integral using the \pd approximant.
\newline
\newline
{\it Subroutine \texttt{MULLO} (\texttt{mullo.f})}
\newline
At a given energy, evaluates the resonance (Mullholland) contribution to the ICS from a pole chosen by hand from the list of available poles.
\newline
\newline
{\it Subroutine \texttt{MULL} (\texttt{mull.f})}
\newline
At a given energy, evaluates the resonance (Mullholland) contribution to the ICS from a pole with the real part closest to that of the pole chosen at the previous energy.
\newline
\newline
{\it Subroutine \texttt{ZPADE} (\texttt{zpade.f})}
\newline
Calculates the full \pd approximant.
\newline
\newline
{\it Subroutine \texttt{ZPR} (\texttt{zpr.f})}
\newline
Evaluates the ratio of the two polynomials in the \pd approximant.
\newline
\newline
{\it Subroutine \texttt{ZRES} (\texttt{zres.f})}
\newline
Calculates (part of)  the residue for a chosen Regge pole from the \pd approximant.
\newline
\newline
{\it Subroutine \texttt{FST3} (\texttt{fst3.f})}
\newline
Supplies the integrand for the integral evaluated in \texttt{tcinte.f}.
\newline
\newline
There are two additional utilities.
\newline
\newline
{\it Subroutine \texttt{SKIP} (\texttt{skip.f})}
\newline
Decides which of the input data/energies must be included in the current run.
\newline
\newline
{\it Subroutine \texttt{SUBTR} (\texttt{subtr.f})}
\newline
Subtracts from the ICS the contribution from a given Regge pole trajectory.
All files are located in the \texttt{ICS/src} directory.
%


\subsection{The  $\texttt{QUADPACK}$ library }
$\texttt{QUADPACK}$ is a $\texttt{FORTRAN}$ subroutine package for the numerical computation of definite one-dimensional integrals. It originated from a joint project of R. Piessens and E. de Doncker (Appl. Math. and Progr. Div.- K.U.Leuven, Belgium), C. Ueberhuber (Inst. Fuer Math.- Techn.U.Wien, Austria), and D. Kahaner (Nation. Bur. of Standards- Washington D.C., U.S.A.) \cite{E1}.
\newline (http://www.netlib.org/quadpack/).

Currently one library subroutine, $\texttt{DQAGS}$, is used in the $\texttt{ICS\_Regge}$. The subroutine estimates integrals over finite intervals using an integrator based on globally adaptive interval subdivision in connection with extrapolation \cite{E2} by the Epsilon algorithm \cite{E3}. The subroutine is called from the $\texttt{ICS\_Regge}$ subroutine $\texttt{tcinte}$. 
For users convenience the whole library is available in the package $\texttt{ICS\_Regge}$.  The link to the library is provided in $\texttt{Makefile}$ and $\texttt{Makefile\_UNIX}$ in $\texttt{ICS/src}$.

\section{Running the  \texttt{ICS\_Regge} code}
\subsection {Creating input data}

Three input files are required for running calculations: a parameter file, \texttt{INPUT}, located in \texttt{ICS/input} directory, and two input files for running \texttt{PADE\_II}. The first file is the \texttt{PADE\_II} parameter file, \texttt{param.pade}. It is created on the fly by running the script \texttt{runICS}. The second file or set of files has(ve) to be supplied in \texttt{ICS/input/PADE\_data} directory and it (or they) contain(s) the data to be Pad\'e approximated. The name of directory \texttt{PADE\_data} can be chosen arbitrary and should be specified in the parameter file \texttt{INPUT} before the starting calculations. The names of the input files in the directory \texttt{PADE\_data} are fixed to be $1,2,.. N_E$, each of which contains previously computed values of $S_{\nu' \gets \nu}(E,J)$ for $J=0,1,2, ...J^{max}_i$, for the energy $E_k$, $k=1,2,..N_E$, and the value of the energy itself. 
An example of input file is given in Appendix B and also provided in directory \texttt{ICS/input}. 
\newline
The file \texttt{ICS/input/INPUT} is self-explanatory and describes each input parameter to be 
specified. Please notice, that each input entry appears between colons (:).
We provide input files for all test cases considered under the names  \texttt{ICS/input/INPUT.BOUND},  \texttt{ICS/input/INPUT.META} and  \texttt{ICS/input/INPUT.FH2}.
\newline 
For an example of the parameter file \texttt{INPUT} see Appendix A.
\subsection {Executing \texttt{ICS\_Regge}}
The script \texttt{runICS} in \texttt{ICS/} directory automates calculations. The following assumptions are made in the script:
\newline
*  all binaries for \texttt{ICS\_Regge} are placed in \texttt{ICS/bin} whereas the binaries for the \texttt{PADE\_II} package are located in \texttt{ICS/PADE/bin}.
\newline
*  input files are located in directory \texttt{ICS/input}. The names of the input files are chosen as described in section 4.1.
\newline
*  output files can be found in \texttt{ICS/output} on completion of the calculation.
We recommend running a calculation in directory \texttt{ICS/}. The command
\texttt{./runICS}
immediately starts the calculation.

\subsection {Understanding the run script \texttt{runICS}}
The run script \texttt{runICS} located in \texttt{ICS/} directory does not require any tuning, editing or corrections in order to start the calculation. Provided that the parameter file \texttt{ICS/input/INPUT} is prepared for calculations, the run script \texttt{runICS} takes care
of the following steps in the following order:
\newline
1. INITIALIZATION
\newline
* Edits input parameter file \texttt{INPUT}
\newline
* Reads input parameters from \texttt{INPUT}
\newline
* Prepares directories for runs
\newline
* Sets the useful directories
\newline
* Cleans the directories if necessary
\newline
2. BUILDING PACKAGES
\newline
* \texttt{PADE\_II}
\newline
* \texttt{ICS\_Regge}
\newline
* Utilities
\newline
3. RUNNING \texttt{ICS\_Regge} FOR ALL INPUT FILES OF INTEREST
\newline
* Checks if the input file falls in the range of energies under investigation
\newline
* Runs \texttt{PADE\_II} with the current input file if it is in the considered range
\newline
* Runs \texttt{ICS\_Regge} if the current input file is in the considered range
\newline
*evaluates the non-resonance background by subtraction the Mulholland contributions from the exact ICS

4. OUTPUT DATA MANAGEMENT
\newline
* Stores the calculated data in the appropriate files

\subsection {Using the code}

Using the code involves at least two steps.
\newline

\subsubsection {Step I}
In the parameter file \texttt{input/INPUT} (see Appendix A) one sets 
\newline
\texttt{is this the first run? :yes:}. The code evaluates the poles $P_i$ and the zeroes $Z_i$ of the \pd approximant (\ref{10}) for each collisional energy $E_k$, for the chosen set of files (see \#5-\#6 of Appendix A), in a region of the CAM plane, 
$ \texttt{x\_min} \le Re P_i, Z_i \le \texttt{x\_max}$, $\texttt{y\_min}\le Im P_i, Z_i \le \texttt{y\_max}$, with the values \texttt{x\_min}, \texttt{x\_max}, \texttt{y\_min}, and \texttt{y\_max} specified by the user in the file \texttt{input/INPUT} (see \#12 -\#15 in the Appendix A). 
\newline
One has the option of not including in the \pd approximant (\ref{10}) the {\it Froissart doublets}, i.e.,   pole-zero pairs, with the distance $|P_i-Z_j| <  \texttt{$\epsilon$}$, with the value of \texttt{$\epsilon$} set in \#11 of Appendix A. Such pairs often represent non-analytical noise present in the input data \cite{G2}, and their removal may be beneficial.
\newline
Also, at this stage the program evaluates, for all energies, the exact ICS using Eq.(\ref{1}) or (\ref{2}), and the first integral in (\ref{6}). The results are written in the files \texttt{output/ics.exact} and \texttt{output/ics.int}, respectively. Provided the energy is entered in $meV$, and the reduced mass is in the $unified\q atomic\q mass\q units$, ($u.a.u$ or $Daltons$) (see \#10 of Appendix A), the cross sections are in the units of angstroms squared
 (\AA$^2$).
 \newline
One then identifies Regge trajectories by plotting the pole positions vs. energy from the output  file  \texttt{output/ics.pole}. (It is recommended to to use the plot $Re P(E)$ vs. $E$, as the real parts of the pole positions are less sensitive to numerical noise).
In the plot the trajectories appear as continuous strings of poles, with additional poles scattered around them in a random manner. 
\newline
\subsubsection {Step II}
In the file \texttt{input/INPUT} one sets \texttt{is this the first run? :no:}. The code takes the first of the files in the energy range \texttt{E\_min}$\le E\le$\texttt{E\_max} with \texttt{E\_min} and \texttt{E\_max} specified by the user in \#8-\#9 of the file \texttt{input/INPUT} (see Appendix A). It then displays all the poles at this energy within the specified range, from which the user chooses the one lying on the Regge trajectory of interest. 
 \newline
Then, if one sets
\texttt{follow trajectory by hand? :no:}the code will follow the trajectory automatically, choosing at the next energy the pole whose real part is closest to that of the pole chosen at the previous energy.
 \newline
If one chooses \texttt{follow trajectory by hand? :yes:} the program continues display the poles,
from which the user must choose the desired one for all values of energy by hand. 
(The "by hand" option is useful, e.g., when working with a poorly defined trajectory from each some of the poles may be missing.)
 \newline
In both cases, the corresponding Mulholland contribution in Eq. (\ref{6}),
\begin{eqnarray}\label{14}
I_n^{Mull}(E)= \frac{8\pi^2}{k_\nu^2}Im \frac{\lambda_n Res[S_{\nu' \leftarrow \nu}(E,\lambda_n)]S^*_{\nu' \leftarrow \nu}(E,\lambda_n^*)}{1+\exp(-2i\pi\lambda_n)}
 \end{eqnarray} 
is written down in the file \texttt{output/ics.mull}.
 \newline
 The pole position and the corresponding residue (in $meV$) are stored in the files \texttt{output/ics.traj} and \texttt{output/ics.resid}.
  \newline
Once the energy \texttt{E\_max} is reached, the program stops and $I_n^{Mull}(E)$ is subtracted from the exact ICS in the file \texttt{output/ics.exact}, the result written in \texttt{output/ics.smooth}.
\newline
Step II can be repeated several times, thus making the program follow different Regge trajectories,
while choosing \texttt{E\_min} and \texttt{E\_max} as is convenient. At the end, the Mulholland contributions from all trajectories considered are subtracted from the exact ICS, and the much less structured non-resonance part of the ICS is stored in the file \texttt{output/ics.smooth}.
\newline
Prior to operating the code one must specify whether the transition is elastic or not, 
by answering \texttt{elastic channel?} with either \texttt{:yes:} or \texttt{:no:}. For single channel scattering 
one may also choose to use the modified Mulholland formula (\ref{9}) by setting in the \texttt{input/INPUT}:
 \texttt{modified  Mulholland? :yes:}
 In this case the Mulholland contribution stored in  \texttt{output/ics.mull} is given by the pole term in
 (\ref{9}), 
\begin{eqnarray}\label{14a}
I_n^{Mull}(E)=  \frac{8\pi^2}{k^2}Im \frac{\lambda_n Res[|S
 (E,\lambda_n)]}{1+\exp(2i\pi\lambda_n)}.
 \end{eqnarray} 
 Note that to use the modified formula (\ref{9}) after using the one given in Eq.(\ref{6}), one must also repeat the step I first. This is necessary to recalculate the exact ICS, from which the contribution (\ref{14}) will then be subtracted.
\section{Examples of using \texttt{ICS\_Regge} }
Three examples of using the $\texttt{ICS\_Regge}$ package are provided. 
The input and output files for these examples are included in the package.
\subsection{Example 1: The hard sphere model (Regge trajectory of the type I)}
The first example 
involves the $S$-matrix element  for potential (single channel) scattering off a hard sphere of a radius $R-d$ surrounded by a thin semi-transparent layer of a radius $R$. 
The spherically symmetric potential $V(r)$ is infinite for $r<R-d$, has a rectangular well of a depth $V$ for $R-d<r<R$, a zero range barrier $\Omega \delta(r-R)$ ($\delta(x)$ is the Dirac delta), and vanishes elsewhere \cite{PLA}. In this example the energy of a non-relativistic particle of a mass $\mu=1\q u.a.u$ varies from $ 1meV$ to $ 100meV$, the radii of the hard sphere and the width of the well $d$  are $2.045$\AA\q and $0.592$\AA, respectively, $V=165$ meV, and
 $\Omega = 1.023\q meV\cdot$ \AA. In this range, there is a single resonance Regge trajectory, originating at $J=0$ in the bound state of the well at about $ - 14 meV$. 
A detailed discussion of this model can be found in Refs.\cite{PLA} where the Regge trajectories were obtained by direct integration of the radial Schroedinger equation for complex values of  $J$.
Here we seek to repeat the results of \cite{PLA} by evaluating the $S$ matrix element for integer $J$'s, and then using the \pd reconstruction.
The data files are in the directory  \texttt{input/BOUND}. 
\newline 
\underline {\it Step I } 
\newline
In the directory  \texttt{input}, copy the file  \texttt{INPUT.BOUND} into the file  \texttt{INPUT}.
Run the code to completion. Use the file  \texttt{output/ics.pole} to plot real parts of the poles vs. energy, and identify the Regge trajectory of interest. Use the file  \texttt{output/ics.exact} to plot the integral cross section in the specified range of energies.
\newline 
\underline {\it Step II } 
\newline
In the file \texttt{INPUT} change \texttt{is this the first run? :yes:} to  \texttt{is this the first run? :no:}.
In the entry \#5 set \texttt{modified  Mulholland? :no:} or \texttt{modified  Mulholland? :yes:}
to use Eq.(\ref{6}) or Eq.(\ref{9}), respectively.
 Run the code. When prompted to choose a pole, 
choose the one at the beginning of the trajectory, with $ReJ \approx 4.85$ and $Im J \approx 0.0043$. After completion, use the files in the directory  \texttt{output} to plot the results. The correct results for the Regge  trajectory, the residues of the resonance pole, the Mulhollolland contribution, and the background cross section are shown in Fig.1.
\begin{figure}[!ht]
\includegraphics[angle=0,width=9 cm]{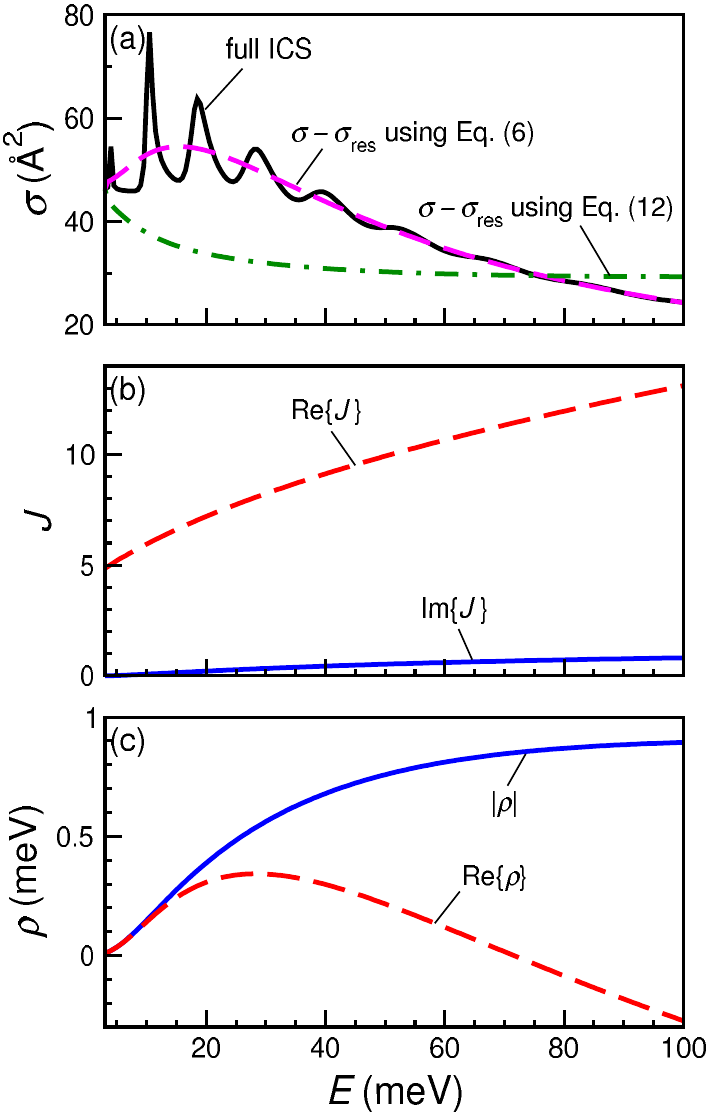}
\caption{{\bf Example 1:} (a) The full ICS (solid) and the smooth background obtained by subtracting from it the resonance contribution $\sigma_{res}$ given by the pole term in Eq.(\ref{4}) (dashed) and in Eq.(\ref{9}) (dot-dashed);
(b) the real (dashed) and the imaginary (solid) parts of the resonance Regge trajectory which affects the ICS in the shown energy range;
(c) the modulus (solid) and the real part (dashed) of the residue of the Regge pole which traces the trajectory in (b).}
\end{figure}
\subsection{Example 2: The hard sphere model (Regge trajectory of the type II)}
This is the same model as in Example 1, but  with  $\Omega = 66.463\q meV\cdot \AA$, considered in the range of collision energies from $ 40meV$ to $ 100meV$. In this case, there is a single resonance Regge trajectory, originating at $J=0$ in a metastable state with the real part of about $48meV$. 
The data files are in the directory  \texttt{input/META}. 
\newline 
\underline {\it Step I } 
\newline
In the directory \texttt{input} copy the file  \texttt{INPUT.META} into the file  \texttt{INPUT}.
Run the code to completion. Use the file  \texttt{output/ics.pole} to plot real parts of the poles vs. energy, and identify the relevant Regge trajectory. Use the file  \texttt{output/ics.exact} to plot the integral cross section in the specified range of energies
\newline 
\underline {\it Step II } 
\newline
In the file  \texttt{INPUT}, change \texttt{is this the first run? :yes:} to  \texttt{is this the first run? :no:}.
In the entry \#5 set \texttt{modified  Mulholland? :no:} or \texttt{modified  Mulholland? :yes:}
to use Eq.(\ref{6}) or Eq.(\ref{9}), respectively. Run the code. When prompted to choose a pole, 
choose the one at the beginning of the trajectory, with $ReJ \approx 0.195$ and $Im J \approx 3.22$. After completion, use the files in the directory  \texttt{output} to plot the results. The correct results for the Regge  trajectory, the residues of the resonance pole, the Mulholland contribution, and the background cross section are shown in Fig. 2.
\begin{figure}[!ht]
\includegraphics[angle=0,width=9 cm]{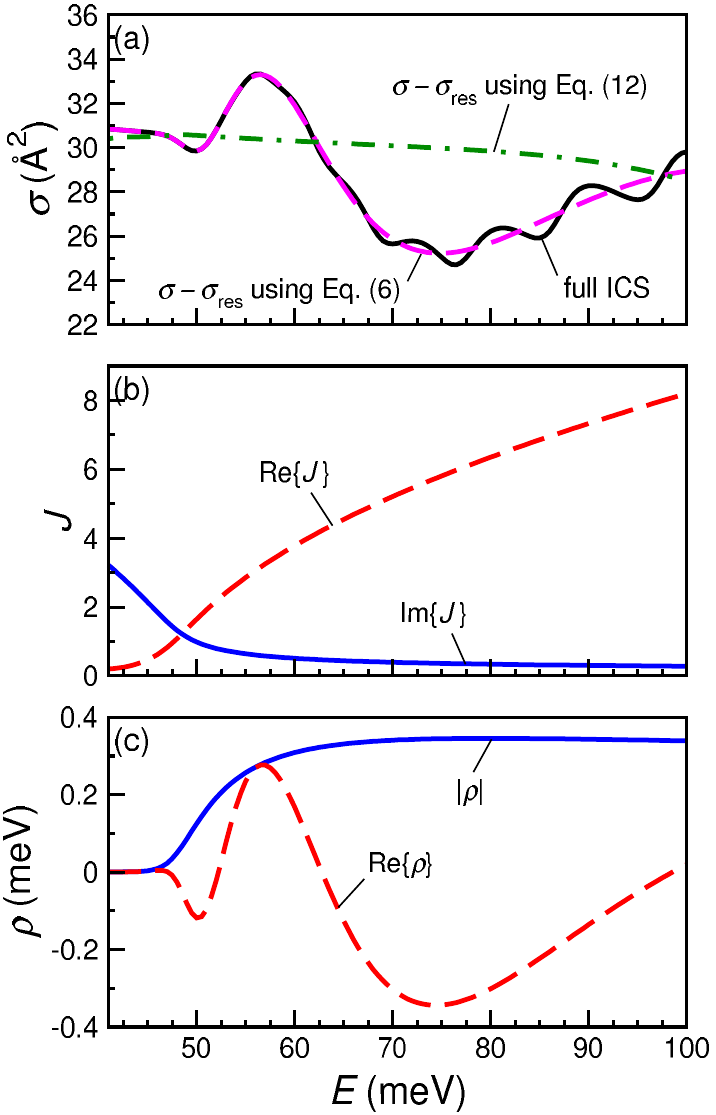}
\caption{{\bf Example 2:} (a) The full ICS (solid) and the smooth background obtained by subtracting from it the resonance contribution $\sigma_{res}$ given by the pole term in Eq.(\ref{4}) (dashed) and in Eq.(\ref{9}) (dot-dashed);
(b) the real (dashed) and the imaginary (solid) parts of the resonance Regge trajectory which affects the ICS in the shown energy range;
(c) the modulus (solid) and the real part (dashed) of the residue of the Regge pole which traces the trajectory in (b).}
\end{figure}
\subsection{Example 3: The $F+H_2(v=0,j= 0,\Omega=0) \to H F+H(v'=2, j'=0,\Omega'=0)$ reaction.  (Two pseudo-crossing Regge trajectories)}
This example uses realistic numerical data obtained in Ref. \cite{DARIO04}, and analysed in Refs. \cite{FH3}-\cite {FH5}, \cite{ICS3}. In the specified energy range there are two resonance Regge trajectories, both contributing to the Regge oscillations seen in the state-to-state integral cross section. At the collision energy of about $38meV$ the imaginary parts of the trajectories cross, while the real parts do not (for details see Ref.\cite{FH4}) The data files are in the directory 
\texttt{input/FH\_2}.
\newline 
\underline {\it Step I } 
\newline
In the directory \texttt{input} copy the file \texttt{INPUT.FH2} into the file \texttt{INPUT}.
Run the code to completion. Use the file \texttt{output/ics.pole} to plot real parts of the poles vs. energy, and identify two relevant Regge trajectories. Use the file \texttt{output/ics.exact} to plot the integral cross section in the specified range of energies.
\newline 
\underline {\it Step II } 
\newline
In the file \texttt{INPUT}, change \texttt{is this the first run? :yes:} to  \texttt{is this the first run? :no:}.
In the entry \#5 keep \texttt{modified  Mulholland? :no:}. 
\newline
respectively. 
Run the code. When prompted to choose a pole, 
choose the one at the beginning of first trajectory, with $ReJ \approx 1.54$ and $Im J \approx 1.17$.
After completion, if necessary, save the data in files \texttt{output/ics.traj}, \texttt{output/ics.mull} and \texttt{output/ics.resid},
as they will be overwritten.
\newline
Run the code again. 
When prompted to choose a pole, this time 
choose the one at the beginning of second trajectory, with $ReJ \approx 5.54$ and $Im J \approx 1.43$.
After completion, use the files in the directory \texttt{output} to plot the results. The correct results for the Regge  trajectories, the residues of the resonance poles, their respective Mulholland contributions, and the background cross section are shown in Fig.3.
\begin{figure}[!ht]
\includegraphics[angle=0,width=9. cm]{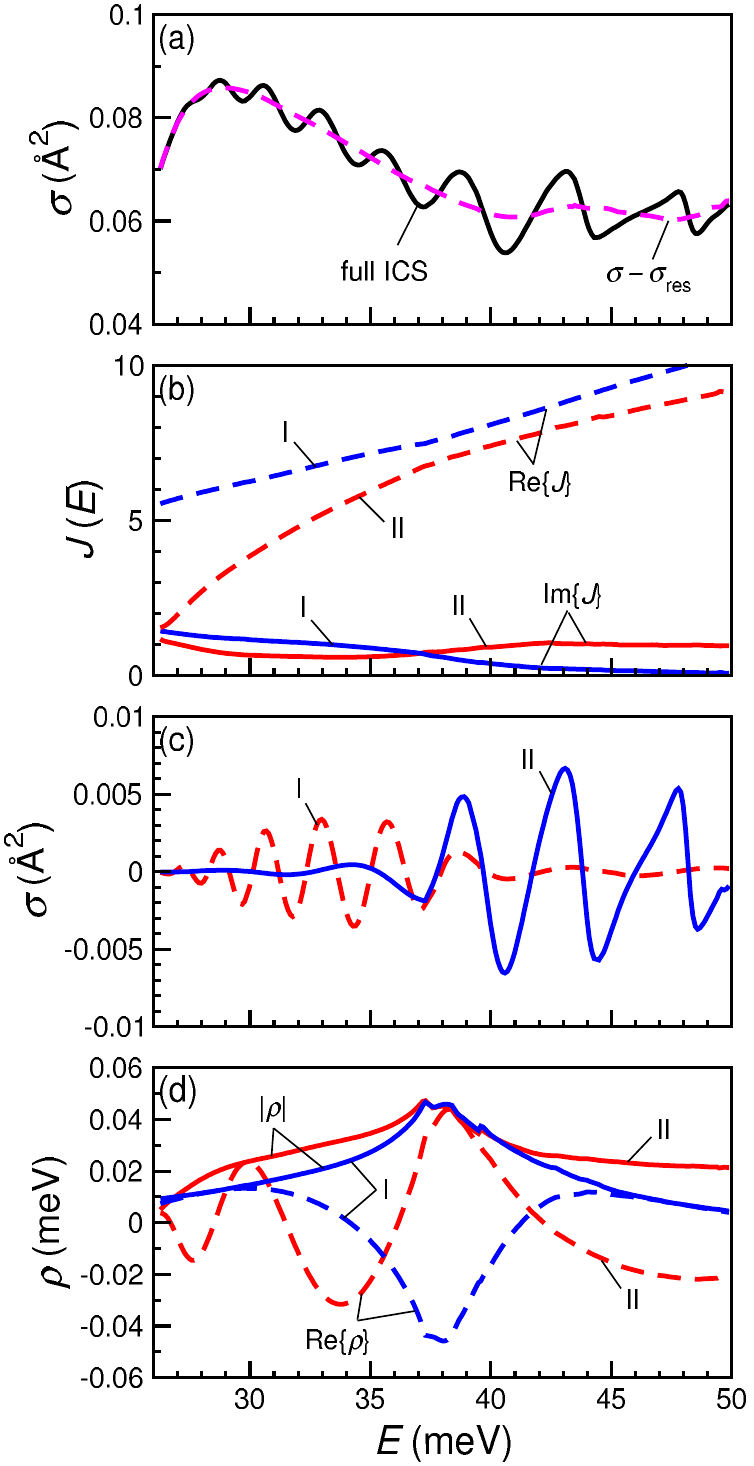}
\caption{{\bf Example 3:} (a) The full ICS (solid) and the smooth background obtained by subtracting from it the resonance contribution $\sigma_{res}$ given by the pole sum term in Eq.(\ref{4}) (dashed);
(b) the real (dashed) and the imaginary (solid) parts of the two resonance Regge trajectories which affect the ICS in the shown energy range;
(c) the Mulholland contributions [pole terms in Eq.(\ref{4})] for the trajectories shown in (b): (I) (solid) and (II) (dashed). 
(d) the moduli (solid) and the real parts (dashed) of the residues of the Regge poles which trace  trajectories (I) and (II) in (b).}
\end{figure}
\newpage
\section{Summary}
In summary, we present a user friendly computer code which evaluates the contribution a resonance Regge trajectory makes to an integral cross section. Regge poles positions and residues are evaluated from numerical values of the corresponding scattering matrix element by Pad\'{e}\q reconstruction. 
The code can be used for analysing elastic, inelastic and reactive transitions. 
\section{Acknowledgements:}
We acknowledge support of the Basque Government (Grant No. IT-472-10), and the Ministry of Science and Innovation of Spain (Grant No. FIS2009-12773-C02-01). The SGI/IZO-SGIker UPV/EHU is acknowledged for providing computational resources.
\newpage
\section{Appendix A: Example of  \texttt{ICS\_Regge} parameter file \texttt{INPUT}} 
\begin{figure}[h!t]
\begin{center}
\includegraphics[angle=0,width=14.5 cm]{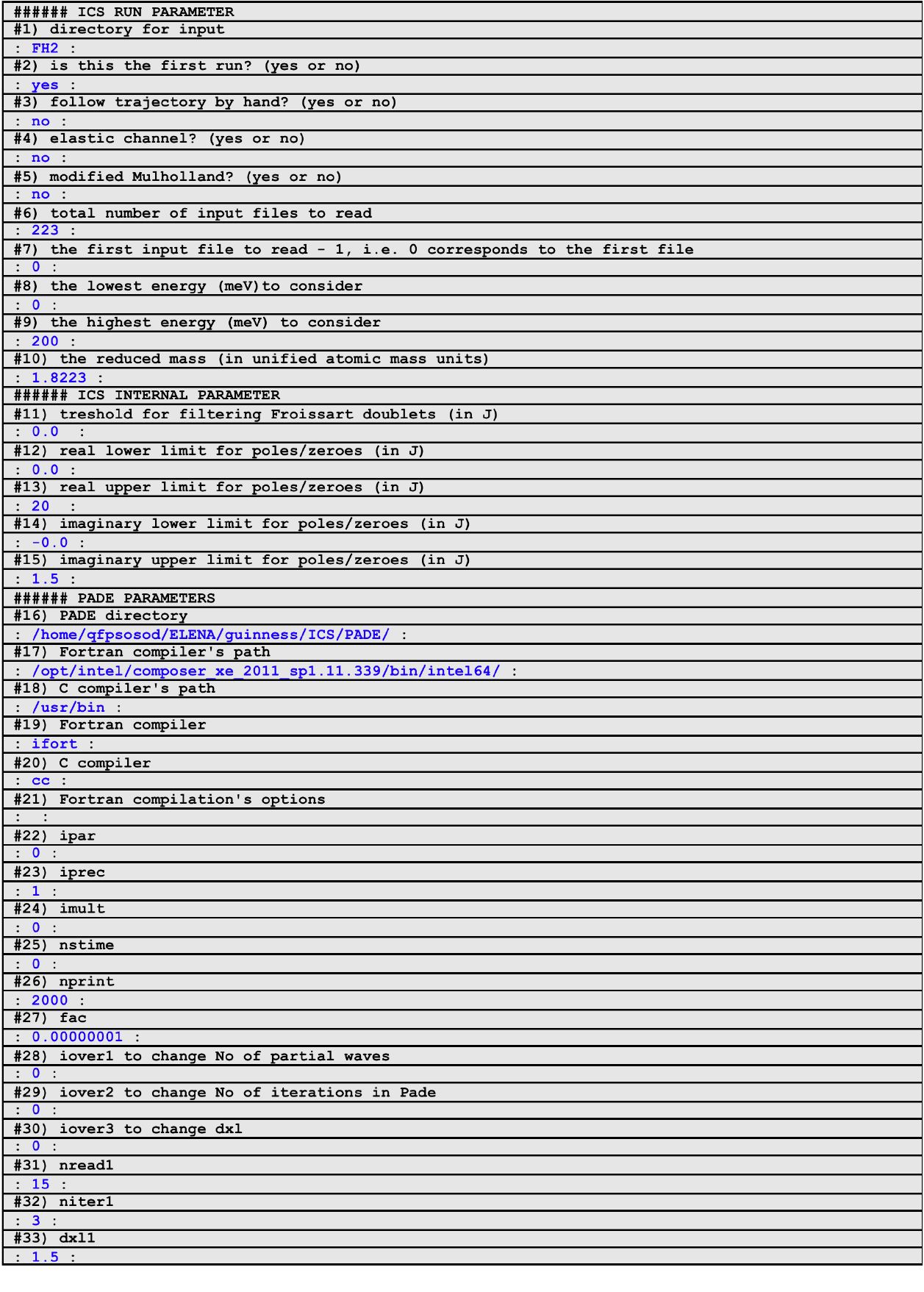}
\end{center}
\caption{An example of the \texttt{INPUT} file for an inelastic or reactive transition.}
\end{figure}
\newpage
\section{Appendix B: Example of  \texttt{ICS\_Regge} input file  \texttt{1}} 

The contents of the file, also described in \cite{PADE2}, include:
\newline
\x {nread}: the number of partial waves read,
\newline
\x{niter}: the number of iterations to remove the quadratic phase,
\newline
\x{sht}: this shifts the input grid points and may be used to avoid exponentiation of extremely large number when evaluating the polynomials involved. The value \x {sht} = \texttt{nread}$/2$ is suggested for a large number of partial waves. 
\newline
\x{jstart} and  \x{jfin}: with all input points numbered by \texttt{j} between 1 and N, determine a range \texttt{jstart} $\le j \le$  \texttt{jfin} to be used for the \pd reconstruction.
\newline
\x{inv}: set to $-1$, not used in present calcualtions, 
\newline
\x{dxl}: determines the width of the strip in which poles and zeroes are removed while evaluating the quadratic phase in Eq. (\ref{9}).

\begin{figure}[!h]
\begin{center}
\includegraphics[angle=0,width=6 cm]{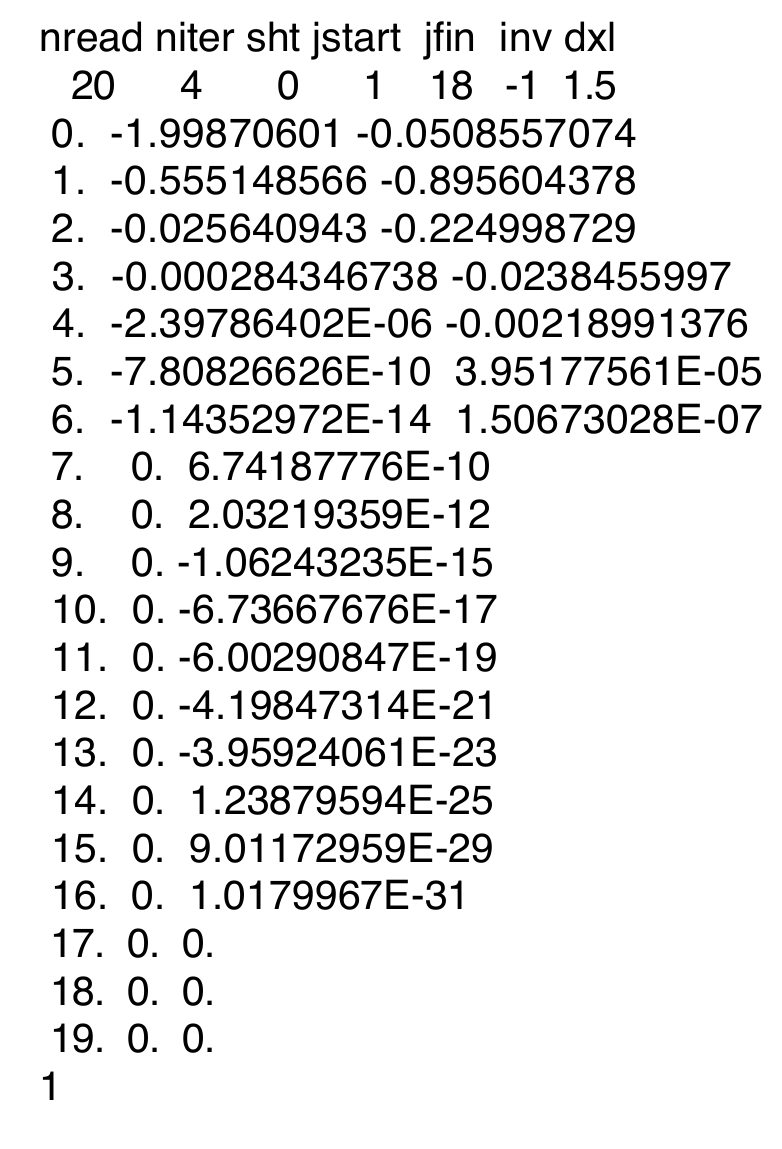}
\end{center}
\caption{The input file \texttt{1}, which contains the input parameters required to run  \texttt{\PD\_II}, the values of the $S$-matrix element for different values of $J$, and the value of collision energy in $meV$. }
\end{figure}

\newpage
\section{Appendix C: Structure of the \texttt{ICS} directory}
\begin{figure}[!ht]
\includegraphics[angle=0,width=15 cm]{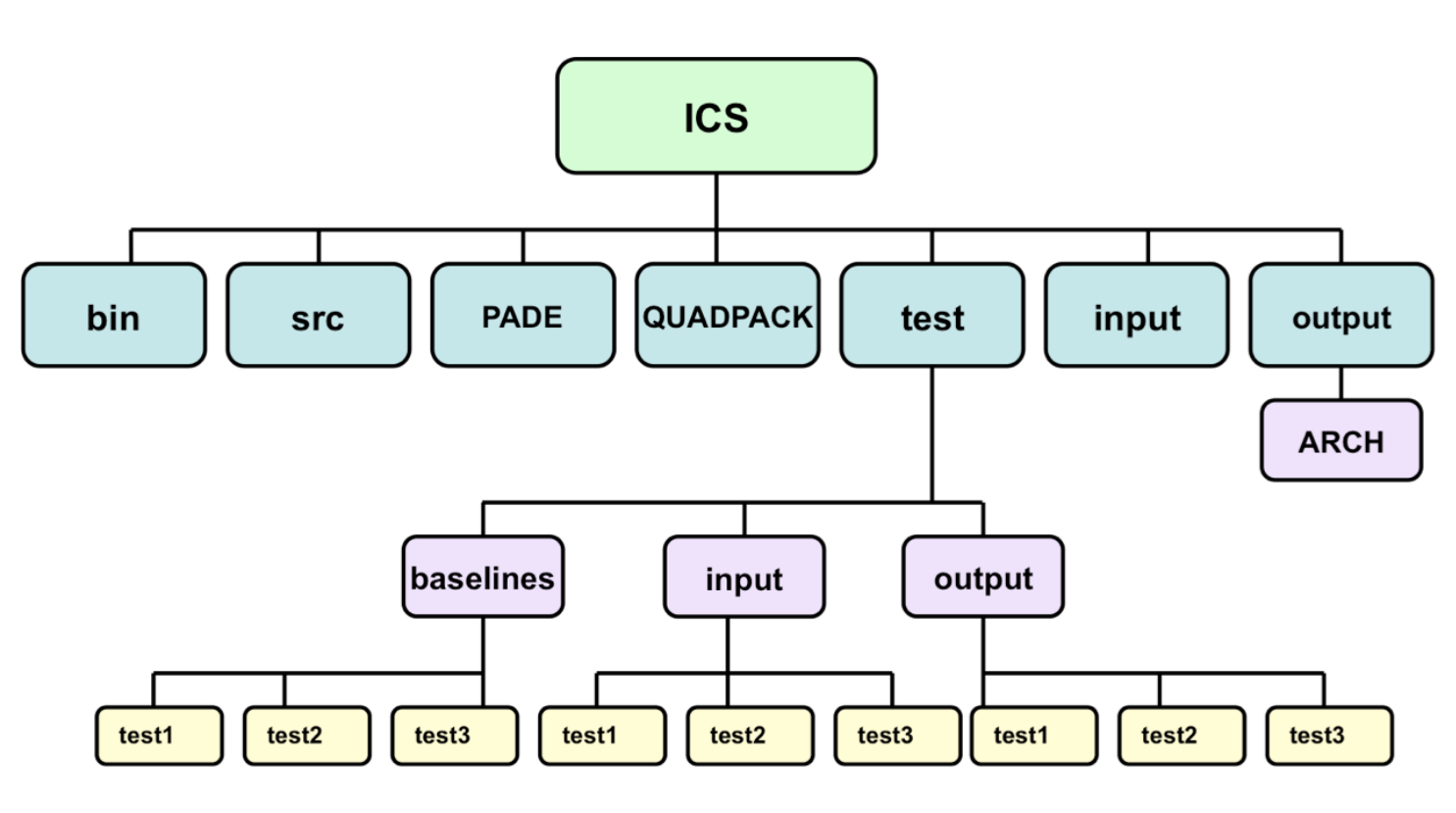}
\caption{Detailed structure of the  \texttt{ICS} directory.}
\end{figure}
\newpage
\section{Appendix D: Structure of the \texttt{PADE} directory}
\begin{figure}[!ht]
\includegraphics[angle=0,width=15 cm]{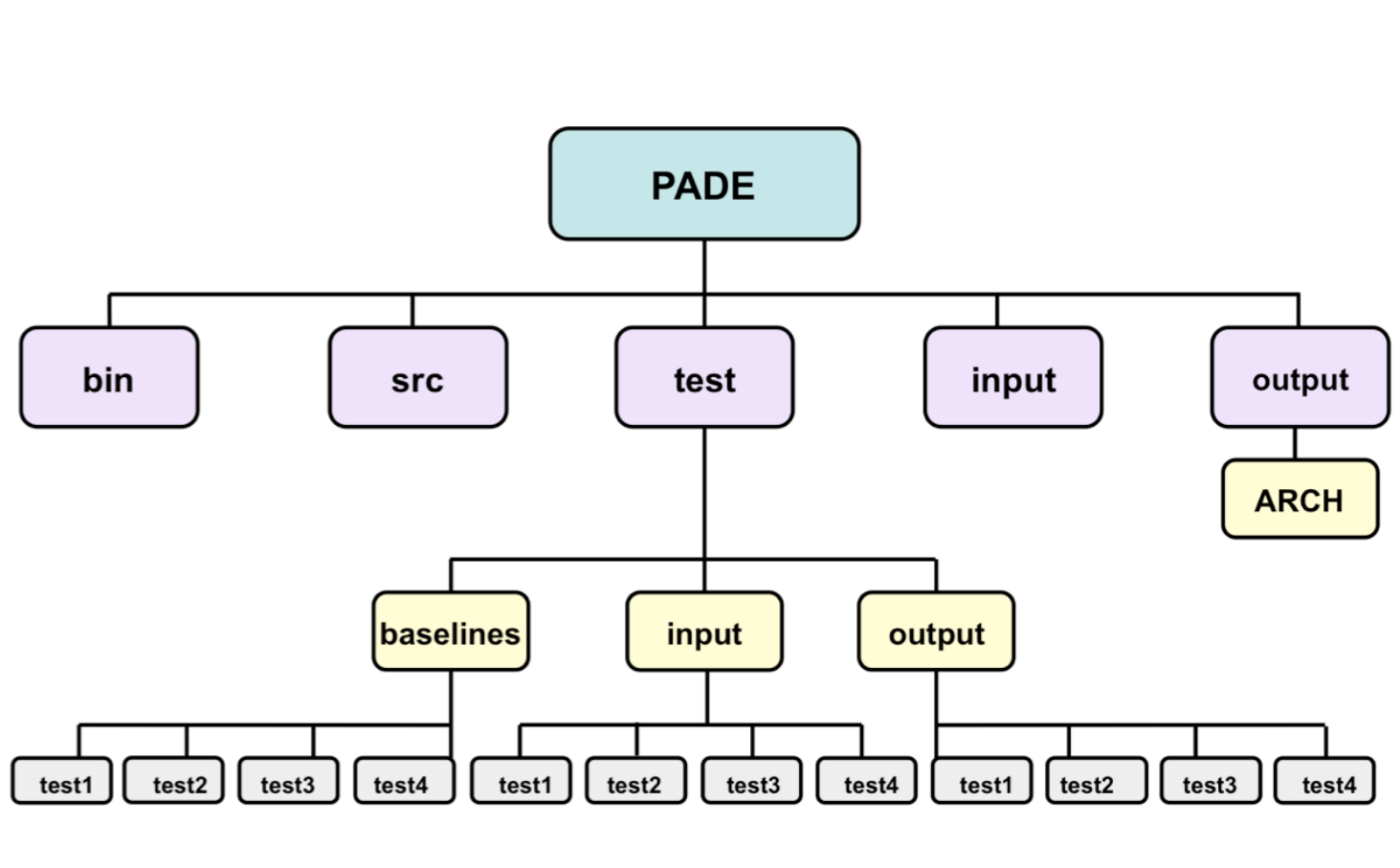}
\caption{Detailed structure of the  \texttt{PADE} directory.}
\end{figure}
\newpage
\section{Appendix E: Structure of the \texttt{QUADPACK} directory}
\begin{figure}[!ht]
\includegraphics[angle=0,width=15 cm]{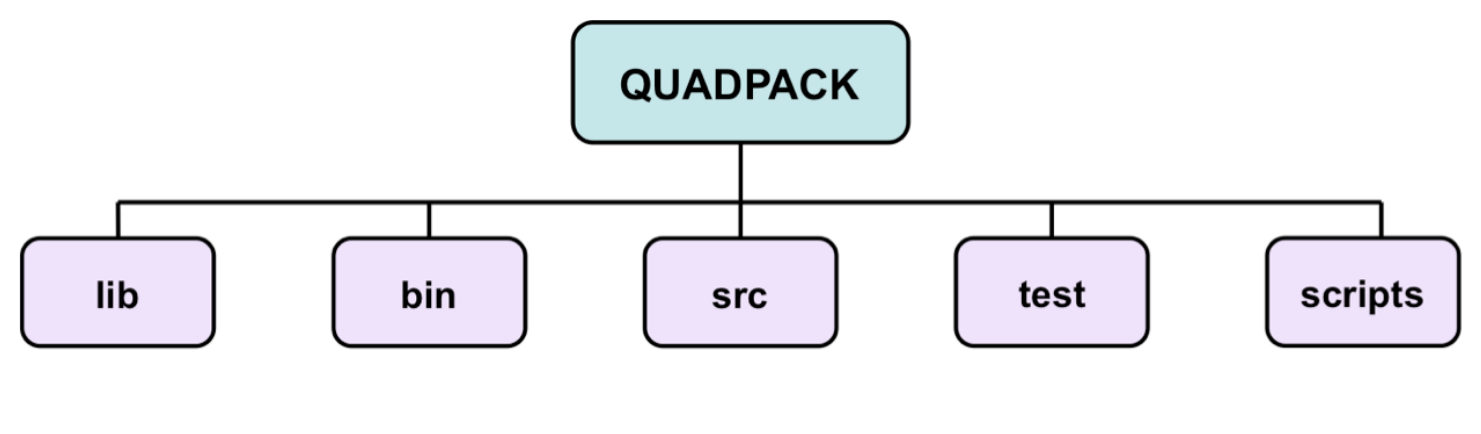}
\caption{Detailed structure of the  \texttt{QUADPACK} directory.}
\end{figure}

\end{document}